\documentclass[11pt,a4paper]{article}
\usepackage{authblk}
\usepackage[hyperref]{acl2019}
\usepackage{times}
\usepackage{latexsym}
\usepackage{graphicx}
\usepackage{multirow}
\usepackage{url}
\usepackage{booktabs}

\aclfinalcopy 
\def\aclpaperid{91} 

\title{UW-BHI at MEDIQA 2019: An Analysis of Representation Methods for Medical Natural Language Inference}

\author[$\clubsuit$]{\textbf{William R. Kearns}}
\author[$\clubsuit$]{\textbf{Wilson Lau}}
\author[$\clubsuit$]{\textbf{Jason A. Thomas \vspace{-10pt}}}
\affil[$\clubsuit$]{Department of Biomedical Informatics and Medical Education, University of Washington}
\affil[ ]{850 Republican Street}
\affil[ ]{Seattle, WA}
\affil[ ]{\fontfamily{courier}\selectfont\{kearnsw, wlau, thomasjt\}@uw.edu}

\date{2019-05-15}

\begin{document}
\maketitle
\begin{abstract}
Recent advances in distributed language modeling have led to large performance increases on a variety of natural language processing (NLP) tasks. However, it is not well understood how these methods may be augmented by knowledge-based approaches. This paper compares the performance and internal representation of an Enhanced Sequential Inference Model (ESIM) between three experimental conditions based on the representation method: Bidirectional Encoder Representations from Transformers (BERT), Embeddings of Semantic Predications (ESP), or Cui2Vec. The methods were evaluated on the Medical Natural Language Inference (MedNLI) subtask of the MEDIQA 2019 shared task. This task relied heavily on semantic understanding and thus served as a suitable evaluation set for the comparison of these representation methods.
\end{abstract}


\section{Introduction}
This paper describes our approach to the Natural Language Inference (NLI) subtask of the MEDIQA 2019 shared task \cite{MEDIQA2019}. As it is not yet clear the extent to which knowledge-based embeddings may provide task-specific improvement over recent advances in contextual embeddings, we provide an analysis of the differences in performance between these two methods. Additionally, it is not yet clear from the literature the extent to which information stored in contextual embeddings overlaps with that in knowledge-based embeddings for which we provide a preliminary analysis of the attention weights of models that use these two representation methods as input. We compare  BERT fine-tuned to MIMIC-III \cite{Johnson2016} and PubMed to Embeddings of Semantic Predications (ESP) trained on SemMedDB and a baseline that uses Cui2Vec embeddings trained on clinical and biomedical text.

Two recent advances in the unsupervised modeling of natural language, Embeddings of Language Models (ELMo) \cite{Peters} and Bidirectional Encoder Representations from Transformers (BERT) \cite{Devlin2018}, have led to drastic improvements across a variety of shared tasks. Both of these methods use transfer learning, a method whereby a multi-layered language model is first trained on a large unlabeled corpus. The weights of the model are then frozen and used as input to a task specific model \cite{Peters, Devlin2018, Liu2019}. This method is particularly well-suited for work in the medical domain where datasets tend to be relatively small due to the high cost of expert annotation. 

However, whereas clinical free-text is difficult to access and share in bulk due to privacy concerns, the biomedical domain is characterized by a significant amount of manually-curated structured knowledge bases. The BioPortal repository currently hosts 773 different biomedical ontologies comprised of over 9.4 million classes. SemMedDB is a triple store that consists of over 94 million predications extracted from PubMed by SemRep, a semantic parser for biomedical text \cite{SemRep, Kilicoglu2012}. These available resources make a strong case for the evaluation of knowledge-based methods for the Medical Natural Language Inference (MedNLI) task \cite{romanov2018}.

\section{Related Work}
In this section, we provide a brief overview of methods for distributional and frame-based semantic representation of natural language. For a more detailed synthesis, we refer the reader to the review of Vector Space Models (VSMs) by Turney and Pantel \shortcite{Turney2010}. 

\subsection{Distributional Semantics}

The distributed representation of words has a long history in computational linguistics, beginning with latent semantic indexing (LSI) \cite{Deerwester1990, Hofmann1999, Kanerva2000RandomIO}, maximum entropy methods \cite{Berger1996}, and latent Dirichlet allocation (LDA) \cite{Blei2003}. More recently, neural network methods have been applied to model natural language \cite{Bengio2000ANP, Weston2008, Ratinov2010}. These methods have been broadly applied as a method of improving supervised model performance by learning word-level features from large unlabeled datasets with more recent work using either Word2Vec \citep{Mikolov, Pavlopoulos2014} or GloVe \cite{Pennington2014} embeddings. Recent work has learned a continuous representation of Unified Medical Language System (UMLS) \cite{Aronson2006} concepts by applying the Word2Vec method to a large corpus of insurance claims, clinical notes, and biomedical text where UMLS concepts were replaced with their Concept Unique Identifiers (CUIs) \cite{Beam2018}.

Models that incorporate sub-word information are particularly useful in the medical domain for representing medical terminology and out-of-vocabulary terms common in clinical notes and consumer health questions \cite{romanov2018}. Most approaches use a temporal convolution over a sliding window of characters and have been shown to improve performance on a variety of tasks \cite{Kim2015, Zhang2015, Minjoon2016, Bojanowski2017EnrichingWV}. 

Embeddings from Language Models (ELMo) computes word representations using a bidirectional language model that consist of a character-level embedding layer followed by a deep bidirectional long short-term memory (LSTM) network \cite{Peters}.
Bidirectional Encoder Representations from Transformers (BERT) replaces the each forward and backward LSTMs with a single Transformer that simultaneously computes attention in both the forward and backward directions and is regarded as the current state-of-the-art method for language representation \cite{Vaswani2017, Devlin2018}. This method additionally substitutes two new unsupervised training objectives in place of the classical language models, i.e., masked language modeling (MLM) and next sentence prediction (NSP). In the case of MLM, a percentage of the words in the corpus are replaced by a [MASK] token. The task is then for the system to predict the masked token. For NSP, the task is given two sentences, $s1$ and $s2$, from a document to determine whether $s2$ is the next sentence following $s1$. 

While ELMo has been shown to outperform GloVe and Word2Vec on consumer health question answering \cite{Kearns2018ResourceAR}, BERT has outperformed ELMo on various clinical tasks \cite{si_enhancing_2019} and has been fine-tuned and applied to the biomedical literature and clinical notes \cite{alsentzer_publicly_2019, huang_clinicalbert:_2019, si_enhancing_2019, lee_biobert:_2019}. BERT supports the transfer of a pretrained general purpose language model to a task-specific application through fine-tuning. The next sentence prediction objective in the pre-training process suggests this method would be inherently suitable for NLI. In addition, BERT utilizes character-based and WordPiece tokenization \cite{wu_googles_2016} to learn the morphological patterns among  inflections. The subword segmentation such as \textit{\#\#nea} in the word \textit{dyspnea} makes it capable to understand the context of an out-of-vocabulary word making it a particularly suitable representation for clinical text.

\subsection{Frame-based Semantics}
FrameNet is a database of sentence-level frame-based semantics that proposes human understanding of natural language is the result of frames in which certain roles are expected to be filled \cite{Baker1997}. For example, the predicate \textit{``replace"} has at least two such roles, the thing being replaced and the new object. 
A sentence such as \textit{``The table was replaced."} raises the question \textit{``With what was the table replaced?"}. Frame-based semantics is a popular approach for semantic role labeling (SRL) \cite{Swayamdipta2018}, question answering (QA) \cite{Shen2007, Roberts2012, He2015, Michael}, and dialog systems \cite{Larsson1998, Gupta2018}.

Vector symbolic architectures (VSA) are an approach that seeks to represent semantic predications by applying binding operators that define a directional transformation between entities \cite{Levy2008}. Early approaches included binary spatter code (BSC) for encoding structured knowledge \cite{Kanerva1996BinarySO, Kanerva1997358FD} and Holographic Embeddings that used circular convolution as a binding operator to improve the scalability of this approach to large knowledge graphs \cite{Plate1995HolographicRR}. The resurgence of neural network methods has focused attention on extending these methods as there is a growing interest in leveraging continuous representations of structured knowledge to improve performance on downstream applications.  

Knowledge graph embeddings (KGE) are one approach that represents entities and their relationships as continuous vectors that are learned using TransE/R \cite{Bordes2009}, RESCAL \cite{Nickel2011ATM}, or Holographic Embeddings \cite{Plate1995HolographicRR, Nickel2015}. Stanovsky et. al \shortcite{Stanovsky2017} showed that RESCAL embeddings pretrained on DbPedia improved performance on the task of adverse drug reaction labeling over a clinical Word2Vec model. RESCAL uses tensor products whose application to representation learning dates back to Smolensky \shortcite{Smolensky1986, Smolensky1990TensorPV} that used the inner product and has recently been applied to the bAbI dataset \cite{Smolensky2016BasicRW, Weston2016TowardsAQ}. Embeddings of Semantic Predications (ESP) are a neural-probabilistic representational approach that uses VSA binding operations to encode structured relationships \cite{Cohen2017}. The Embeddings Augmented by Random Permutations (EARP) used in this paper are a modified ESP approach that applies random permutations to the entity vectors during training and were shown to improve performance on the Bigger Analogy Test Set by up to 8\%   against a fastText baseline \cite{CohenCoNLL2018}.

\section{Methods}
In this section, we provide details on the three representation methods used in this study, i.e. BERT, Cui2Vec, and ESP. We continue with a description of the inference model used in each experiment to predict the label for a given hypothesis/premise pair. 

\begin{figure*}[t!]
    \includegraphics[width=\textwidth, height=11cm]{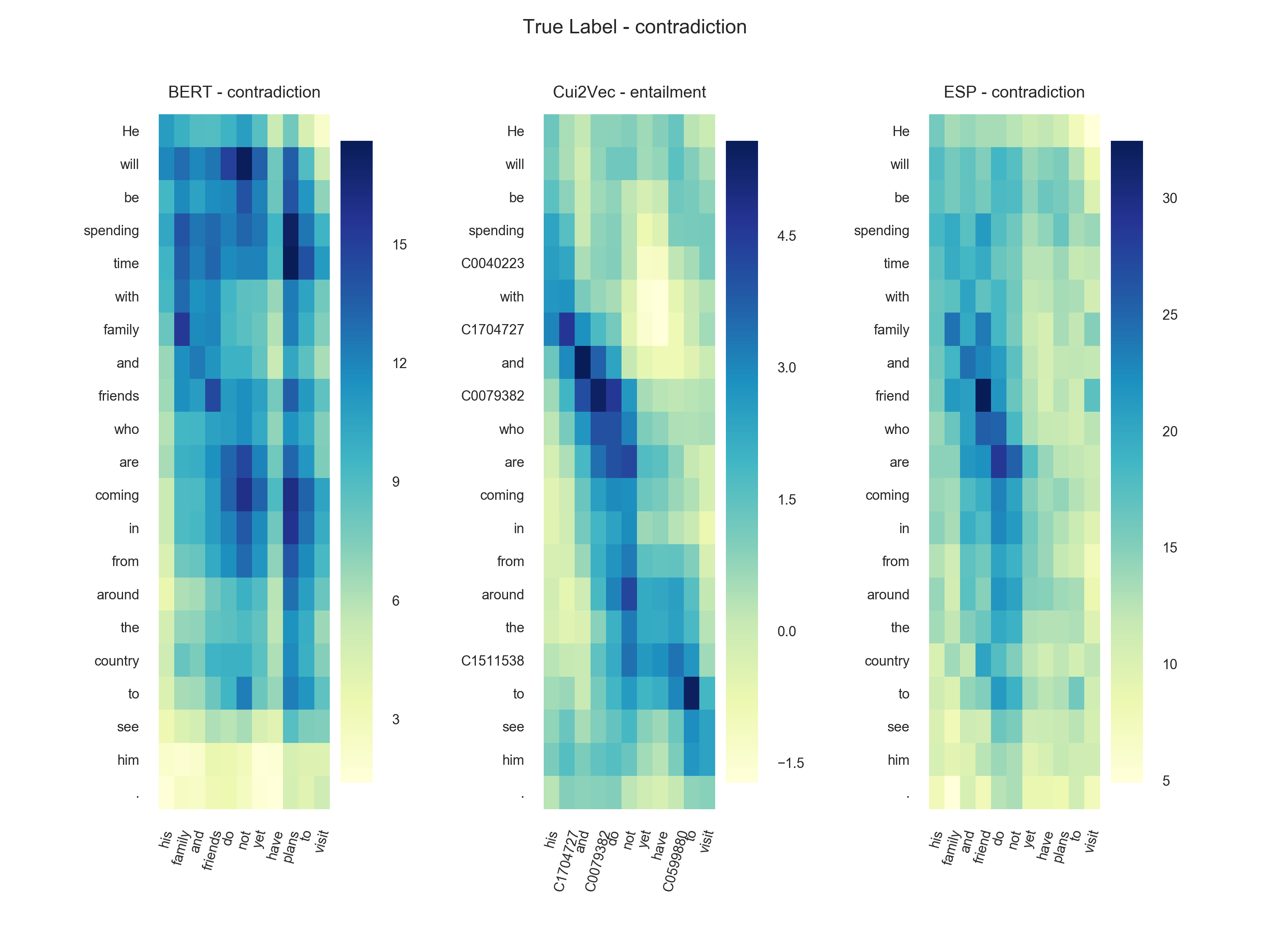}
    \caption{An example of a correct BERT prediction demonstrating its general domain coverage and contextual embedding. Premise: \textit{``He will be spending time with family and friends who are coming in from around the country to see him."} Hypothesis: \textit{``his family and friends do not yet have plans to visit."}} 
    \label{fig:HeatmapTopBert}
\end{figure*}

\begin{figure*}[t!]
    \includegraphics[width=\textwidth,
    height=11cm]{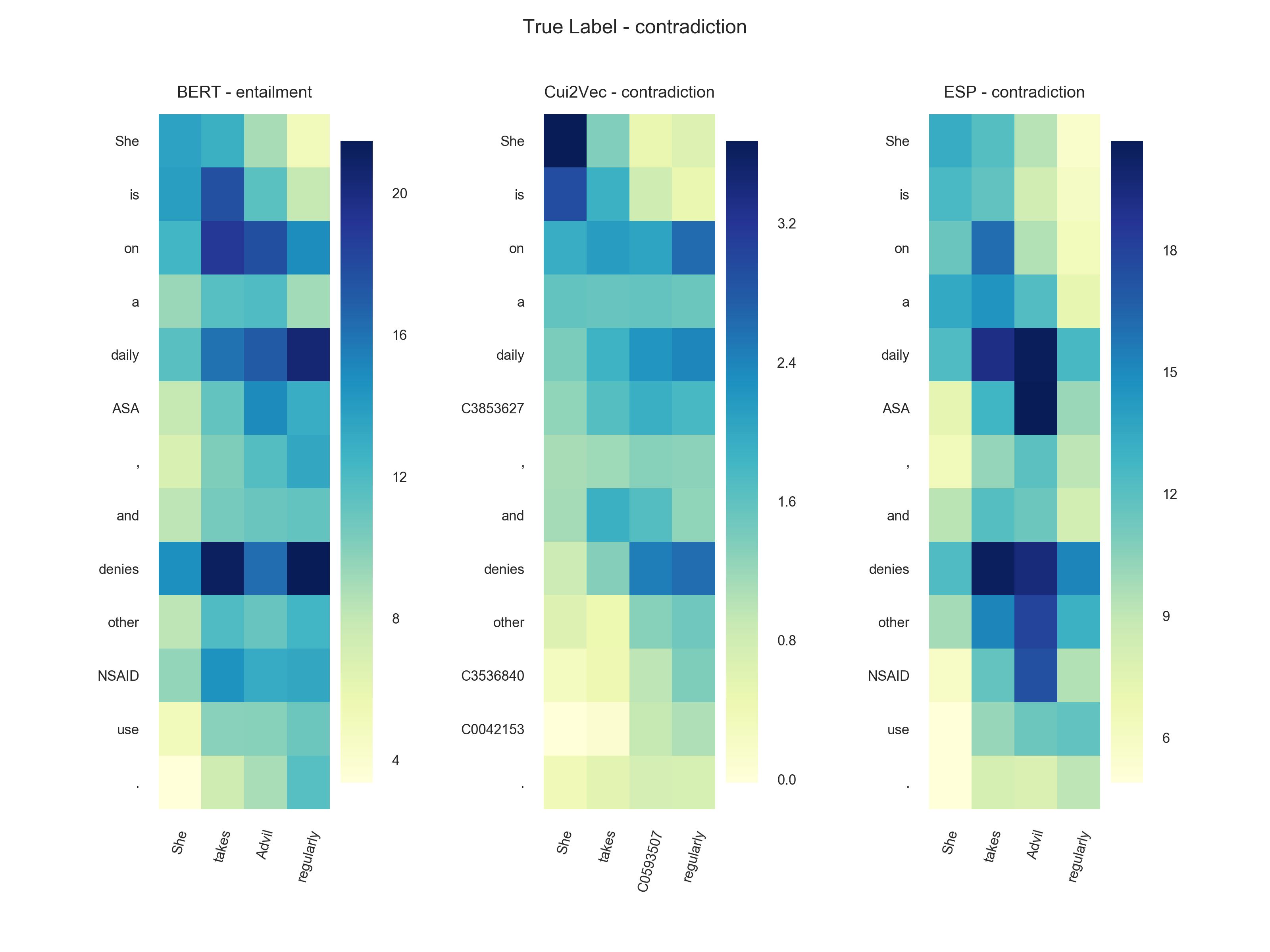}
    \caption{An example of a correct ESP prediction demonstrating its ability to associate Advil as a subclass of NSAIDs. Premise: \textit{``She is on a daily ASA, and denies other NSAID use."} Hypothesis: \textit{``She takes Advil regularly."}} 
    \label{fig:HeatmapTopESP}
\end{figure*}

\subsection{Representation Layer}
There are many publicly available biomedical BERT embeddings which were initialized from the original BERT Base models. BioBERT  was trained on PubMed Abstracts and PubMed Central Full-text articles \cite{lee_biobert:_2019}.  In this study, we applied ClinicalBERT that was initialized from BioBERT and subsequently trained on all MIMIC-III notes \cite{alsentzer_publicly_2019}.

For Cui2Vec, we used the publicly available implementation from Beam et al. \shortcite{Beam2018} that was trained on a corpus consisting of 20 million clinical notes from a research hospital, 1.7 million full-text articles from PubMed, and an insurance claims database with 60 million members.

For ESP, we used a 500-dimensional model trained over SemMedDB using the recent Embeddings Augmented by Random Permutations (EARP) approach with a $10^{-7}$ sampling threshold for predications and a $10^{-5}$ sampling threshold for concepts excluding concepts that had a frequency greater than $10^{6}$ \cite{CohenCoNLL2018}. 

To apply Cui2Vec and ESP, we first processed the MedNLI dataset \cite{romanov2018} with MetaMap to normalize entities to their concept unique identifier (CUI) in the UMLS \cite{Aronson2006}. MetaMap takes text as input and applies biomedical and clinical entity recognition (ER), followed by word sense disambiguation (WSD) that links entities to their normalized concept unique identifiers (CUIs). Entities that mapped to a UMLS CUI were assigned a representation in Cui2Vec and ESP. Other tokens were assigned vector representations using fastText embeddings trained on MIMIC-III data \cite{Bojanowski2017EnrichingWV, romanov2018}.   

\subsection{Inference Model}
For all experiments, we used the AllenNLP implementation \cite{Gardner2017} of the Enhanced Sequential Inference Model (ESIM) architecture \cite{ESIM}. This model encodes the premise and hypothesis using a Bidirectional LSTM (BiLSTM) where at each time step the hidden state of the LSTMs are concatenated to represent its context. Local inference between the two sentences is then achieved by aligning the relevant information between words in the premise and hypothesis. This alignment based on soft attention is implemented by the inner product between the encoded premise and encoded hypothesis to produce an attention matrix (Figure \ref{fig:HeatmapTopBert} and \ref{fig:HeatmapTopESP}). These attention values are used to create a weighted representation of both sentences. An enhanced representation of the premise is created by concatenating the encoded premise, the weighted hypothesis, the encoded premise minus the weighted hypothesis, and the element-wise multiplication of the encoded premise and the weighted hypothesis. The enhanced representation of the hypothesis is created similarly. This operation is expected to enhance the local inference information between elements in each sentence. This representation is then projected into the original dimension and fed into a second BiLSTM inference layer in order to capture inference composition sequentially. The resulting vector is then summarized by max and average pooling. These two pooled representations are concatenated and passed through a multi-layered perceptron followed by a sigmoid function to predict probabilities for each of the sentence labels, i.e. \textit{entailment}, \textit{contradiction}, and \textit{neutral}.

\section{Results}
The ESIM model achieved an accuracy of 81.2\%, 65.2\%, and 77.8\% for the MedNLI task using BERT, Cui2Vec, and ESP, respectively. Table \ref{tab:predictionAccuracyData} shows the number of correct predictions by each embedding type. The BERT model has the highest accuracy on predicting \textit{entailment} and \textit{contradiction} labels, while the ESP model has the highest accuracy on predicting \textit{neutral} labels. However, the difference is only significant in the case of \textit{entailment}. 

\begin{figure*}[t!]
    \includegraphics[width=\textwidth]{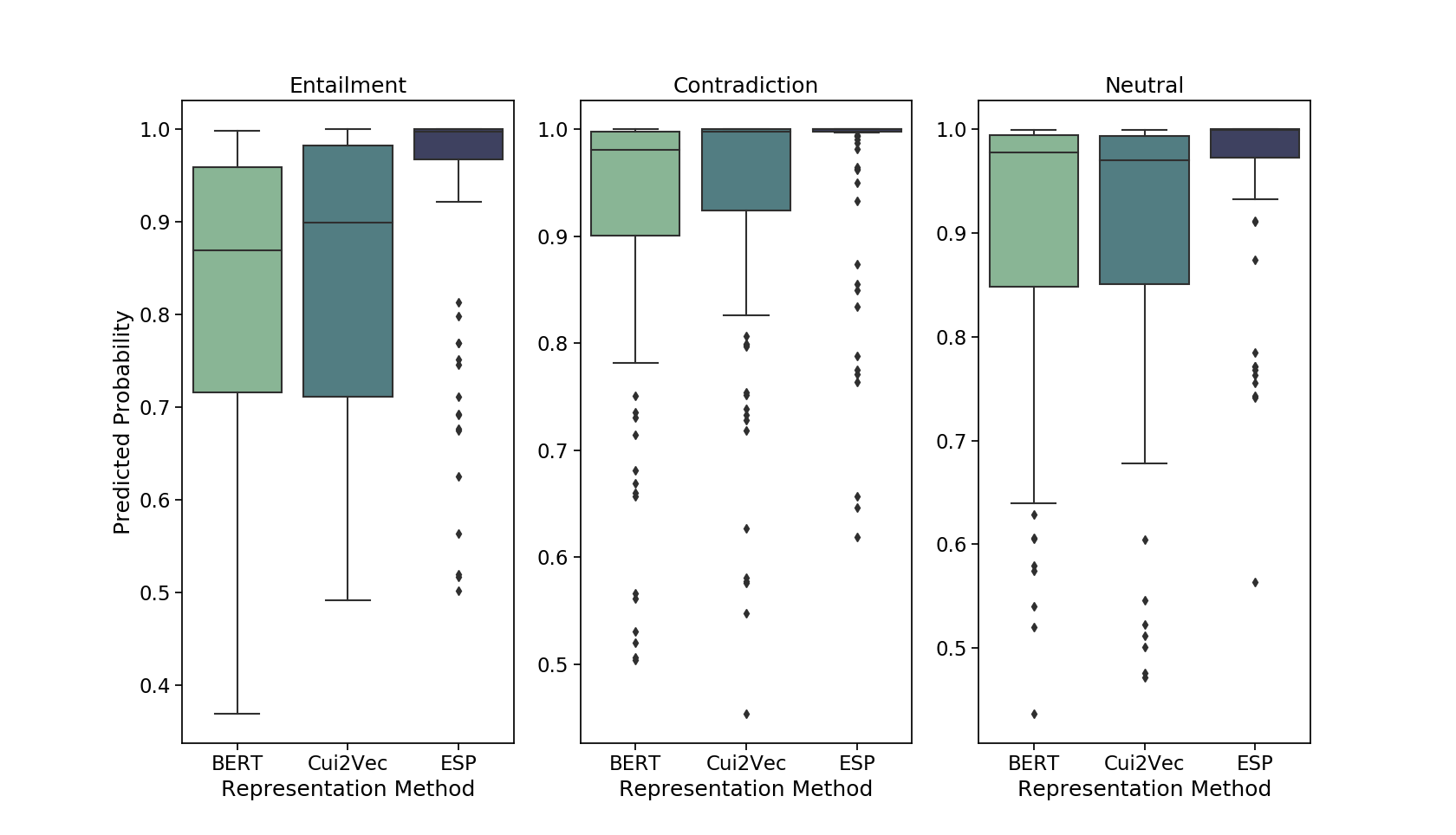}
    \caption{Distribution of predicted probability of the gold label from the subset of correct predictions for each representation method.}
    \label{fig:boxplot}
\end{figure*}

To evaluate the ability to set a predictive threshold for use in clinical applications, we sought to measure the certainty with which the model made its predictions. To achieve this goal, we used the predicted probabilities of each embedding type on their respective subset of correct predictions such that. We found the predicted probability of ESP to be much higher than the others as depicted in Figure \ref{fig:boxplot}. ESP's minimum predicted probability as well as the variance of its distribution is the lowest among all embedding types. 

\begin{table*}[t!]
\begin{tabular}{@{}llll@{}}
  & \multicolumn{3}{c}{\textbf{Embedding Type}} \\ 
  \textbf{Label} &  \textbf{BERT} & \textbf{Cui2Vec} &  \textbf{ESP} \\ 
\hline Entailment & \textbf{82.22\% (n=111)} & 60.00\% (n=81) & 71.85\% (n=97) \\ 
Contraction & \textbf{88.15\% (n=119)} & 74.81\% (n=101) & 87.41\% (n=118)  \\
Neutral &  73.33\% (n=99) & 60.74\% (n=82) & \textbf{74.07\% (n=100)} \\
\end{tabular}
\centering
\caption{Model accuracy for each label by embedding type.}
\label{tab:predictionAccuracyData}
\end{table*}

\subsection{Error Analysis}
To examine the relationship between embedding prediction performance and hypothesis focus, we first annotated the test set for: 
\begin{itemize}
    \item hypothesis focus (e.g. \textit{medications, procedures, symptoms}, etc.)
    \item hypothesis tense (e.g. \textit{past, current, future}) 
\end{itemize} 

\subsubsection{Focus}
A total of eleven, non-mutually exclusive hypothesis focus classes were arrived at by consensus of the three authors after an initial blinded round of annotation by two annotators. The remaining data was annotated by one of these annotators. We provide definitions of the classes and their overall counts in Table \ref{table:Foci}. The classes are: \textit{State}, \textit{Anatomy}, \textit{Disease}, \textit{Process}, \textit{Temporal}, \textit{Medication}, \textit{Clinical Finding}, \textit{Location}, \textit{Lab/Imaging}, \textit{Procedure}, and \textit{Examination}. 

We then performed Pearson's chi-squared test with Yates' continuity correction on 2x2 contingency tables for each embedding sentence pair prediction (correct or incorrect) with each hypothesis focus (presence or absence) using the \textit{chisq.test} function in R software and results reported in Table \ref{table:chiSquared}.
 
The only significant relationships between hypothesis focus and embedding accuracy were found between BERT and \textit{Disease} (p-value = 0.01) and Cui2Vec and \textit{Disease} (p-value = 0.01) through Pearson's Chi-squared test with Yates' continuity correction. 
Both embeddings achieved higher accuracy on sentence pairs with a hypothesis focus labeled \textit{Disease} (BERT=90.4\%; Cui2Vec=76.6\%) than without (BERT=78.5\%; Cui2Vec=61.7\%).

\begin{table*}[t!]
\begin{tabular}{lll}
\hline \textbf{Hypothesis Focus} & \textbf{Definition} & \textbf{Count(\%)} \\ \hline
State & Patient state or symptoms (e.g. \textit{``...has high blood pressure..."}) & 251 (62.0) \\
Anatomy & Specific body part referenced (e.g. \textit{``... has back pain"}) & 115 (28.4) \\
Disease & Similar to state, but a defined disease (e.g. \textit{``...has Diabetes"}) & 95 (23.5) \\
\multirow{2}{*}{Process} & Events like transfers, family visiting, scheduling, or vague & 52 (12.8) \\
& references to interventions (e.g. \textit{``...received medical attention"}) &\\
\multirow{2}{*}{Temporal} & Reference to time (e.g. \textit{``...initial blood pressure was low"}) & 51 (12.6)\\
& besides tense or history & \\
\multirow{2}{*}{Medication} & Any reference to medication (e.g. \textit{``antibiotics", ``fluids"}, & 32 (7.9) \\
& \textit{``oxygen", ``IV"}) including administration and patient habits & \\
Clinical Finding & Results of an exam, lab/image, procedure, or a diagnosis & 28 (6.9)\\
Location & Specific physical location specified (e.g.\textit{``...discharged home"}) & 28 (6.9) \\
Lab/Imaging & Laboratory tests or imaging (e.g. \textit{histology, CBC, CT scan}) & 24 (5.9) \\
\multirow{2}{*}{Procedure} & Physical procedure besides Lab/Image or exam & 14 (3.5)  \\
& (e.g.\textit{ ``intubation", ``surgery", ``biopsies"}) & \\
Examination & Physical examination or explicit use of the word exam(ination) & 3 (0.7) 
\end{tabular}
\caption{\label{font-table} Hypothesis foci definitions, examples, and count for all 405 hypotheses in the test set.}
\label{table:Foci}
\centering
\end{table*}

\begin{table*}[t!]
\begin{center}
\begin{tabular}{@{}llllllllll@{}}
  & \multicolumn{9}{c}{\textbf{Embedding Type}} \\
  & \multicolumn{3}{c}{\textbf{BERT}} & \multicolumn{3}{c}{\textbf{Cui2Vec}} & \multicolumn{3}{c}{\textbf{ESP}} \\
 \cmidrule(lr){2-4}\cmidrule(lr){5-7}\cmidrule(lr){8-10}
 \textbf{Focus} & \textbf{(+)} & \textbf{(-)} & \textbf{p-value} & \textbf{(+)} & \textbf{(-)} & \textbf{p-value} & \textbf{(+)} & \textbf{(-)} & \textbf{p-Value} \\
\hline Anatomy & 93 & 22 & 1 & 73 & 42 & 0.74 & 90 & 25 & 0.99 \\
Clinical Finding & 24 & 4 & 0.71 & 16 & 12 & 0.47 & 24 & 4 & 0.42  \\
Disease & 85 & 9 & \textbf{0.01} & 72 & 22 & \textbf{0.01} & 78 & 16 & 0.21 \\
Examination & 3 & 0 & 0.93 & 2 & 1 & 0.58 & 3 & 0 & 0.82 \\
Lab/Imaging & 30 & 7 & 1 & 22 & 15 & 0.55 & 31 & 6 & 0.48 \\
Location & 21 & 7 & 0.53 & 14 & 14 & 0.12 & 19 & 9 & 0.28 \\
Medication & 27 & 5 & 0.81 & 24 & 8 & 0.30 & 28 & 4 & 0.25 \\
Procedure & 12 & 2 & 0.93 & 7 & 7 & 0.35 & 11 & 3 & 1 \\
Process & 41 & 11 & 0.78 & 35 & 17 & 0.85 & 40 & 12 & 1 \\
State & 198 & 53 & 0.16 & 158 & 93 & 0.27 & 191 & 60 & 0.36 \\
Temporal & 38 & 12 & 0.41 & 37 & 13 & 0.22 & 41 & 9 & 0.56
\end{tabular}
\caption{Results from chi-squared (with Yates' continuity correction) test of correct(+) and incorrect(-) predictions by embedding and hypothesis focus type.}
\label{table:chiSquared}
\centering
\end{center}
\end{table*}

\subsubsection{Tense}
Each hypothesis was annotated for tense into one of three mutually exclusive classes: \textit{Past, Current,} and \textit{Future}. 
Test set hypotheses were predominantly \textit{Current} (n=273; 67.4\%) or \textit{Past} (n=131; 32.3\%) tense. 
Only one hypothesis (0.2\%) was \textit{Future} tense. 
A subset (n=22; 7.9\%) of the \textit{Current} tense hypotheses explicitly described patient history (e.g. \textit{``The patient has a history of PE"}).

\section{Discussion}
Our preliminary analysis, identified several patterns from the attention heatmaps that differentiated the three representation methods. We describe two here and provide the entire set of attention matrices along with supplemental analysis on Github \footnote{\url{https://kearnsw.github.io/MEDIQA-2019/}}. 

The coverage of entities and their associations was characteristic of BERT predictions (Figure \ref{fig:HeatmapTopBert}). BERT associated \textit{``spending time"} with \textit{``plans"} in addition to the lexical overlap of the word \textit{``family"} which is attended by each experimental condition in this example. All three embeddings identified the contradictory significance of the word \textit{``not"} in the hypothesis. However, BERT associated it with both spans \textit{``will be"} and \textit{``are coming"} in the premise, which led to the correct prediction. Cui2Vec over-attended the lexical match of the words \textit{``and"}, \textit{``to"} and \textit{``C0079382"}, which led to the wrong prediction.

The ESP model recognized hierarchical relationships between entities, e.g.  \textit{``Advil"} and \textit{``NSAIDs"} (Figure \ref{fig:HeatmapTopESP}). In this example, the ESP approach attends to the daily use of \textit{``ASA"} (acetyl-salicylic acid), i.e. aspirin, and the patient denying the use of \textit{``other NSAIDs"}. This pattern was recognized multiple times in our analysis and provides a strong example of how continuous representations of biomedical ontologies may be used to augment contextual representations. 

\section{Limitations}
The results presented in this paper compare a single model for each representation method fine-tuned to the development set. However, it is well known that the weights of the same model may vary slightly between training runs. Therefore, a more comprehensive approach would be to present the average attention weights across multiple training runs and to examine the weights at each attention layer of the models which we leave for future work. 

\section{Conclusion}
We have presented our analysis of representation methods on the MedNLI task as evaluated during the MEDIQA 2019 shared task. We found that BERT embeddings fine-tuned using PubMed and MIMIC-III outperformed both Cui2Vec and ESP methods.  However, we found that ESP had the lowest variance and highest predictive certainty, which may be useful in determining a minimum threshold for clinical decision support systems. \textit{Disease} was the only hypothesis focus to show a significant positive relationship with embedding prediction accuracy. This association was present for BERT and Cui2Vec embeddings - but not ESP. Overall, contradiction was the easiest label to predict for all three embeddings, which may be the result of an annotation artifact where contradiction pairs had higher lexical overlap often differentiated by explicit negation. However, overfitting on the negation can lead to lower accuracy on other entailment labels. Further, our preliminary results indicate that recognition of hierarchical relationships is characteristic of ESP suggesting that they can be used to augment contextual embeddings which, in turn, would contribute lexical coverage including sub-word information. We propose combining these methods in future work.

\section*{Acknowledgments}

We would like to acknowledge Trevor Cohen for sharing the Embeddings of Semantic Predications used in this study. Author Jason A. Thomas' work was supported, in part, by the National Library of Medicine (NLM) Training Grant T15LM007442. This work was facilitated, in part, through the use of the advanced computational, storage, and networking infrastructure managed by the Research Computing Club at the University of Washington and funded by an STF award. \\
 
\bibliography{acl2019}

\begin{thebibliography}{57}
\expandafter\ifx\csname natexlab\endcsname\relax\def\natexlab#1{#1}\fi

\bibitem[{Alsentzer et~al.(2019)Alsentzer, Murphy, Boag, Weng, Jin, Naumann,
  and McDermott}]{alsentzer_publicly_2019}
Emily Alsentzer, John~R. Murphy, Willie Boag, Wei{-}Hung Weng, Di~Jin, Tristan
  Naumann, and Matthew B.~A. McDermott. 2019.
\newblock \href {http://arxiv.org/abs/1904.03323} {Publicly available clinical
  {BERT} embeddings}.
\newblock \emph{CoRR}, abs/1904.03323.

\bibitem[{Aronson(2006)}]{Aronson2006}
Alan~R Aronson. 2006.
\newblock \href
  {http://0-skr.nlm.nih.gov.library.law.suffolk.edu/papers/references/metamap06.pdf}
  {{Metamap: Mapping text to the umls metathesaurus}}.
\newblock \emph{Bethesda MD NLM NIH DHHS}, pages 1--26.

\bibitem[{Baker et~al.(1998)Baker, Fillmore, and Lowe}]{Baker1997}
Collin~F. Baker, Charles~J. Fillmore, and John~B. Lowe. 1998.
\newblock \href {https://doi.org/10.3115/980845.980860} {The berkeley framenet
  project}.
\newblock In \emph{Proceedings of the 36th Annual Meeting of the Association
  for Computational Linguistics and 17th International Conference on
  Computational Linguistics - Volume 1}, ACL '98/COLING '98, pages 86--90,
  Stroudsburg, PA, USA. Association for Computational Linguistics.
\newblock \url{https://doi.org/10.3115/980845.980860}.

\bibitem[{Beam et~al.(2018)Beam, Kompa, Fried, Palmer, Shi, Cai, and
  Kohane}]{Beam2018}
Andrew~L. Beam, Benjamin Kompa, Inbar Fried, Nathan~P. Palmer, Xu~Shi, Tianxi
  Cai, and Isaac~S. Kohane. 2018.
\newblock \href {http://arxiv.org/abs/1804.01486} {Clinical concept embeddings
  learned from massive sources of medical data}.
\newblock \emph{CoRR}, abs/1804.01486.

\bibitem[{{Ben Abacha} et~al.(2019){Ben Abacha}, Shivade, and
  Demner-Fushman}]{MEDIQA2019}
Asma {Ben Abacha}, Chaitanya Shivade, and Dina Demner-Fushman. 2019.
\newblock Overview of the mediqa 2019 shared task on textual inference,
  question entailment and question answering.
\newblock In \emph{Proceedings of the BioNLP 2019 workshop, Florence, Italy,
  August 1, 2019}. Association for Computational Linguistics.

\bibitem[{Bengio et~al.(2003)Bengio, Ducharme, Vincent, and
  Janvin}]{Bengio2000ANP}
Yoshua Bengio, R{\'e}jean Ducharme, Pascal Vincent, and Christian Janvin. 2003.
\newblock \href
  {http://dl.acm.org.offcampus.lib.washington.edu/citation.cfm?id=944919.944966}
  {A neural probabilistic language model}.
\newblock \emph{J. Mach. Learn. Res.}, 3:1137--1155.

\bibitem[{Berger et~al.(1996)Berger, Pietra, and Pietra}]{Berger1996}
Adam~L. Berger, Vincent J.~Della Pietra, and Stephen A.~Della Pietra. 1996.
\newblock \href {http://dl.acm.org/citation.cfm?id=234285.234289} {A maximum
  entropy approach to natural language processing}.
\newblock \emph{Comput. Linguist.}, 22(1):39--71.

\bibitem[{Blei et~al.(2003)Blei, Ng, and Jordan}]{Blei2003}
David~M. Blei, Andrew~Y. Ng, and Michael~I. Jordan. 2003.
\newblock \href {http://dl.acm.org/citation.cfm?id=944919.944937} {Latent
  dirichlet allocation}.
\newblock \emph{J. Mach. Learn. Res.}, 3:993--1022.

\bibitem[{Bojanowski et~al.(2017)Bojanowski, Grave, Joulin, and
  Mikolov}]{Bojanowski2017EnrichingWV}
Piotr Bojanowski, Edouard Grave, Armand Joulin, and Tomas Mikolov. 2017.
\newblock \href {https://doi.org/10.1162/tacl_a_00051} {Enriching word vectors
  with subword information}.
\newblock \emph{Transactions of the Association for Computational Linguistics},
  5:135--146.
\newblock \url{https://doi.org/10.1162/tacl\_a\_00051}.

\bibitem[{Bordes and Weston(2009)}]{Bordes2009}
Antoine Bordes and Jason Weston. 2009.
\newblock \href {https://doi.org/10.1016/j.procs.2017.05.045} {{Learning
  Structured Embeddings of Knowledge Bases}}.
\newblock \emph{Artificial Intelligence}, (Bengio):301--306.
\newblock \url{https://doi.org/10.1016/j.procs.2017.05.045}.

\bibitem[{Chen et~al.(2017)Chen, Zhu, Ling, Wei, Jiang, and Inkpen}]{ESIM}
Qian Chen, Xiaodan Zhu, Zhen-Hua Ling, Si~Wei, Hui Jiang, and Diana Inkpen.
  2017.
\newblock \href {https://doi.org/10.18653/v1/P17-1152} {Enhanced {LSTM} for
  natural language inference}.
\newblock In \emph{Proceedings of the 55th Annual Meeting of the Association
  for Computational Linguistics (Volume 1: Long Papers)}, pages 1657--1668,
  Vancouver, Canada. Association for Computational Linguistics.
\newblock \url{https://doi.org/10.18653/v1/P17-1152}.

\bibitem[{Cohen and Widdows(2017)}]{Cohen2017}
Trevor Cohen and Dominic Widdows. 2017.
\newblock \href {https://doi.org/10.1016/j.jbi.2017.03.003} {{Embedding of
  semantic predications}}.
\newblock \emph{Journal of Biomedical Informatics}, 68:150--166.
\newblock \url{https://doi.org/10.1016/j.jbi.2017.03.003}.

\bibitem[{Cohen and Widdows(2018)}]{CohenCoNLL2018}
Trevor Cohen and Dominic Widdows. 2018.
\newblock \href {https://www.aclweb.org/anthology/K18-1045} {Bringing order to
  neural word embeddings with embeddings augmented by random permutations
  ({EARP})}.
\newblock In \emph{Proceedings of the 22nd Conference on Computational Natural
  Language Learning}, pages 465--475, Brussels, Belgium. Association for
  Computational Linguistics.

\bibitem[{Deerwester et~al.(1990)Deerwester, Dumais, Landauer, Furnas, and
  Harshman}]{Deerwester1990}
Scott~C. Deerwester, Susan~T. Dumais, Thomas~K. Landauer, George~W. Furnas, and
  Richard~A. Harshman. 1990.
\newblock Indexing by latent semantic analysis.
\newblock \emph{JASIS}, 41:391--407.

\bibitem[{Devlin et~al.(2018)Devlin, Chang, Lee, and Toutanova}]{Devlin2018}
Jacob Devlin, Ming{-}Wei Chang, Kenton Lee, and Kristina Toutanova. 2018.
\newblock \href {http://arxiv.org/abs/1810.04805} {{BERT:} pre-training of deep
  bidirectional transformers for language understanding}.
\newblock \emph{CoRR}, abs/1810.04805.

\bibitem[{Gardner et~al.(2018)Gardner, Grus, Neumann, Tafjord, Dasigi, Liu,
  Peters, Schmitz, and Zettlemoyer}]{Gardner2017}
Matt Gardner, Joel Grus, Mark Neumann, Oyvind Tafjord, Pradeep Dasigi,
  Nelson~F. Liu, Matthew Peters, Michael Schmitz, and Luke Zettlemoyer. 2018.
\newblock \href {https://www.aclweb.org/anthology/W18-2501} {{A}llen{NLP}: A
  deep semantic natural language processing platform}.
\newblock In \emph{Proceedings of Workshop for {NLP} Open Source Software
  ({NLP}-{OSS})}, pages 1--6, Melbourne, Australia. Association for
  Computational Linguistics.

\bibitem[{Gupta et~al.(2018)Gupta, Shah, Mohit, Kumar, and Lewis}]{Gupta2018}
Sonal Gupta, Rushin Shah, Mrinal Mohit, Anuj Kumar, and Mike Lewis. 2018.
\newblock \href {http://arxiv.org/abs/1810.07942} {Semantic parsing for task
  oriented dialog using hierarchical representations}.
\newblock \emph{CoRR}, abs/1810.07942.

\bibitem[{He(2015)}]{He2015}
Luheng He. 2015.
\newblock \href {https://doi.org/10.18653/v1/D15-1076} {{Question-Answer Driven
  Semantic Role Labeling : Using Natural Language to Annotate Natural
  Language}}.
\newblock \emph{Emnlp2015}, (September):643--653.
\newblock \url{https://doi.org/10.18653/v1/D15-1076}.

\bibitem[{Hofmann(1999)}]{Hofmann1999}
Thomas Hofmann. 1999.
\newblock \href {https://doi.org/10.1145/312624.312649} {Probabilistic latent
  semantic indexing}.
\newblock In \emph{Proceedings of the 22Nd Annual International ACM SIGIR
  Conference on Research and Development in Information Retrieval}, SIGIR '99,
  pages 50--57, New York, NY, USA. ACM.
\newblock \url{https://doi.org/10.1145/312624.312649}.

\bibitem[{Huang et~al.(2019)Huang, Altosaar, and
  Ranganath}]{huang_clinicalbert:_2019}
Kexin Huang, Jaan Altosaar, and Rajesh Ranganath. 2019.
\newblock \href {http://arxiv.org/abs/1904.05342} {{ClinicalBERT}: {Modeling}
  {Clinical} {Notes} and {Predicting} {Hospital} {Readmission}}.
\newblock \emph{arXiv:1904.05342 [cs]}.
\newblock ArXiv: 1904.05342.

\bibitem[{Johnson et~al.(2016)Johnson, Pollard, Shen, Lehman, Feng, Ghassemi,
  Moody, Szolovits, Celi, and Mark}]{Johnson2016}
Alistair E~W Johnson, Tom~J Pollard, Lu~Shen, Li-wei~H Lehman, Mengling Feng,
  Mohammad Ghassemi, Benjamin Moody, Peter Szolovits, Leo~Anthony Celi, and
  Roger~G Mark. 2016.
\newblock {MIMIC-III, a freely accessible critical care database}.
\newblock pages 1--9.

\bibitem[{Kanerva(1996)}]{Kanerva1996BinarySO}
Pentti Kanerva. 1996.
\newblock \href {http://dl.acm.org/citation.cfm?id=646256.684603} {Binary
  spatter-coding of ordered k-tuples}.
\newblock In \emph{Proceedings of the 1996 International Conference on
  Artificial Neural Networks}, ICANN 96, pages 869--873, London, UK, UK.
  Springer-Verlag.

\bibitem[{Kanerva(1997)}]{Kanerva1997358FD}
Pentti Kanerva. 1997.
\newblock Fully distributed representation.
\newblock In \emph{In Proceedings Real World Computing Symposium (Report
  TR-96001)}, pages 358--365.

\bibitem[{Kanerva et~al.(2000)Kanerva, Kristoferson, and
  Holst}]{Kanerva2000RandomIO}
Pentti Kanerva, Jan Kristoferson, and Anders Holst. 2000.
\newblock Random indexing of text samples for latent semantic analysis.
\newblock In \emph{In Proceedings of the 22nd Annual Conference of the
  Cognitive Science Society}, pages 103--6. Erlbaum.

\bibitem[{Kearns and Thomas(2018)}]{Kearns2018ResourceAR}
William~R Kearns and Jason~A Thomas. 2018.
\newblock \href {https://www.ncbi.nlm.nih.gov/pmc/articles/PMC6371272/}
  {Resource and response type classification for consumer health question
  answering}.
\newblock \emph{AMIA Annual Symposium proceedings. AMIA Symposium},
  2018:634–--643.

\bibitem[{Kilicoglu et~al.(2012)Kilicoglu, Shin, Fiszman, Rosemblat, and
  Rindflesch}]{Kilicoglu2012}
Halil Kilicoglu, Dongwook Shin, Marcelo Fiszman, Graciela Rosemblat, and
  Thomas~C. Rindflesch. 2012.
\newblock \href {https://doi.org/10.1093/bioinformatics/bts591} {{SemMedDB: A
  PubMed-scale repository of biomedical semantic predications}}.
\newblock \emph{Bioinformatics}, 28(23):3158--3160.
\newblock \url{https://doi.org/10.1093/bioinformatics/bts591}.

\bibitem[{Kim et~al.(2015)Kim, Jernite, Sontag, and Rush}]{Kim2015}
Yoon Kim, Yacine Jernite, David Sontag, and Alexander~M. Rush. 2015.
\newblock \href {http://arxiv.org/abs/1508.06615} {Character-aware neural
  language models}.
\newblock \emph{CoRR}, abs/1508.06615.

\bibitem[{Larsson and Traum(2000)}]{Larsson1998}
Staffan Larsson and David~R. Traum. 2000.
\newblock \href {https://doi.org/10.1017/S1351324900002539} {Information state
  and dialogue management in the trindi dialogue move engine toolkit}.
\newblock \emph{Nat. Lang. Eng.}, 6(3-4):323--340.
\newblock \url{https://doi.org/10.1017/S1351324900002539}.

\bibitem[{Lee et~al.(2019)Lee, Yoon, Kim, Kim, Kim, So, and
  Kang}]{lee_biobert:_2019}
Jinhyuk Lee, Wonjin Yoon, Sungdong Kim, Donghyeon Kim, Sunkyu Kim, Chan~Ho So,
  and Jaewoo Kang. 2019.
\newblock \href {http://arxiv.org/abs/1901.08746} {{BioBERT}: a pre-trained
  biomedical language representation model for biomedical text mining}.
\newblock \emph{arXiv:1901.08746 [cs]}.

\bibitem[{Levy and Gayler(2008)}]{Levy2008}
Simon~D. Levy and Ross Gayler. 2008.
\newblock \href {http://dl.acm.org/citation.cfm?id=1566174.1566215} {Vector
  symbolic architectures: A new building material for artificial general
  intelligence}.
\newblock In \emph{Proceedings of the 2008 Conference on Artificial General
  Intelligence 2008: Proceedings of the First AGI Conference}, pages 414--418,
  Amsterdam, The Netherlands, The Netherlands. IOS Press.

\bibitem[{Liu et~al.(2019)Liu, He, Chen, and Gao}]{Liu2019}
Xiaodong Liu, Pengcheng He, Weizhu Chen, and Jianfeng Gao. 2019.
\newblock Multi-task deep neural networks for natural language understanding.
\newblock \emph{CoRR}, abs/1901.11504.

\bibitem[{Michael et~al.(2018)Michael, Stanovsky, He, Dagan, and
  Zettlemoyer}]{Michael}
Julian Michael, Gabriel Stanovsky, Luheng He, Ido Dagan, and Luke Zettlemoyer.
  2018.
\newblock \href {https://doi.org/10.18653/v1/N18-2089} {Crowdsourcing
  question-answer meaning representations}.
\newblock In \emph{Proceedings of the 2018 Conference of the North {A}merican
  Chapter of the Association for Computational Linguistics: Human Language
  Technologies, Volume 2 (Short Papers)}, pages 560--568, New Orleans,
  Louisiana. Association for Computational Linguistics.
\newblock \url{https://doi.org/10.18653/v1/N18-2089}.

\bibitem[{Mikolov et~al.(2013)Mikolov, Sutskever, Chen, Corrado, and
  Dean}]{Mikolov}
Tomas Mikolov, Ilya Sutskever, Kai Chen, Greg Corrado, and Jeffrey Dean. 2013.
\newblock \href {http://dl.acm.org/citation.cfm?id=2999792.2999959}
  {Distributed representations of words and phrases and their
  compositionality}.
\newblock In \emph{Proceedings of the 26th International Conference on Neural
  Information Processing Systems - Volume 2}, NIPS'13, pages 3111--3119, USA.
  Curran Associates Inc.

\bibitem[{Nickel et~al.(2015)Nickel, Rosasco, and Poggio}]{Nickel2015}
Maximilian Nickel, Lorenzo Rosasco, and Tomaso~A. Poggio. 2015.
\newblock \href {http://arxiv.org/abs/1510.04935} {Holographic embeddings of
  knowledge graphs}.
\newblock \emph{CoRR}, abs/1510.04935.

\bibitem[{Nickel et~al.(2011)Nickel, Tresp, and Kriegel}]{Nickel2011ATM}
Maximilian Nickel, Volker Tresp, and Hans-Peter Kriegel. 2011.
\newblock \href {http://dl.acm.org/citation.cfm?id=3104482.3104584} {A
  three-way model for collective learning on multi-relational data}.
\newblock In \emph{Proceedings of the 28th International Conference on
  International Conference on Machine Learning}, ICML'11, pages 809--816, USA.
  Omnipress.

\bibitem[{Pavlopoulos et~al.(2014)Pavlopoulos, Kosmopoulos, and
  Androutsopoulos}]{Pavlopoulos2014}
Ioannis Pavlopoulos, Aris Kosmopoulos, and Ion Androutsopoulos. 2014.
\newblock {Continuous Space Word Vectors Obtained by Applying Word2Vec to
  Abstracts of Biomedical Articles}.
\newblock \url{http://bioasq.lip6.fr/info/BioASQword2vec/}.

\bibitem[{Pennington et~al.(2014)Pennington, Socher, and
  Manning}]{Pennington2014}
Jeffrey Pennington, Richard Socher, and Christopher Manning. 2014.
\newblock \href {https://doi.org/10.3115/v1/D14-1162} {Glove: Global vectors
  for word representation}.
\newblock In \emph{Proceedings of the 2014 Conference on Empirical Methods in
  Natural Language Processing ({EMNLP})}, pages 1532--1543, Doha, Qatar.
  Association for Computational Linguistics.
\newblock \url{https://doi.org/10.3115/v1/D14-1162}.

\bibitem[{Peters et~al.(2018)Peters, Neumann, Iyyer, Gardner, Clark, Lee, and
  Zettlemoyer}]{Peters}
Matthew~E. Peters, Mark Neumann, Mohit Iyyer, Matt Gardner, Christopher Clark,
  Kenton Lee, and Luke Zettlemoyer. 2018.
\newblock \href {http://arxiv.org/abs/1802.05365} {Deep contextualized word
  representations}.
\newblock \emph{CoRR}, abs/1802.05365.

\bibitem[{{Plate}(1995)}]{Plate1995HolographicRR}
T.~A. {Plate}. 1995.
\newblock \href {https://doi.org/10.1109/72.377968} {Holographic reduced
  representations}.
\newblock \emph{IEEE Transactions on Neural Networks}, 6(3):623--641.
\newblock \url{https://doi.org/10.1109/72.377968}.

\bibitem[{Rindflesch and Fiszman(2003)}]{SemRep}
Thomas~C Rindflesch and Marcelo Fiszman. 2003.
\newblock The interaction of domain knowledge and linguistic structure in
  natural language processing: interpreting hypernymic propositions in
  biomedical text.
\newblock \emph{Journal of Biomedical Informatics}, 36(6):462--477.
\newblock \url{https://doi.org/10.1016/j.jbi.2003.11.003}.

\bibitem[{Roberts and Demner-fushman(2016)}]{Roberts2012}
Kirk Roberts and Dina Demner-fushman. 2016.
\newblock \href {https://www.ncbi.nlm.nih.gov/pmc/articles/PMC5428549/}
  {{Annotating Logical Forms for EHR Questions}}.
\newblock In \emph{Proceedings of the 10th International Conference on Language
  Resources and Evaluation}, Section 3, pages 3772--3778.

\bibitem[{Romanov and Shivade(2018)}]{romanov2018}
Alexey Romanov and Chaitanya Shivade. 2018.
\newblock \href {http://arxiv.org/abs/1808.06752} {Lessons from natural
  language inference in the clinical domain}.
\newblock \emph{CoRR}, abs/1808.06752.

\bibitem[{Seo et~al.(2016)Seo, Kembhavi, Farhadi, and Hajishirzi}]{Minjoon2016}
Min~Joon Seo, Aniruddha Kembhavi, Ali Farhadi, and Hannaneh Hajishirzi. 2016.
\newblock \href {http://arxiv.org/abs/1611.01603} {Bidirectional attention flow
  for machine comprehension}.
\newblock \emph{CoRR}, abs/1611.01603.

\bibitem[{Shen and Lapata(2007)}]{Shen2007}
Dan Shen and Mirella Lapata. 2007.
\newblock \href {http://www.aclweb.org/anthology/D/D07/D07-1002} {Using
  semantic roles to improve question answering}.
\newblock In \emph{Proceedings of the 2007 Joint Conference on Empirical
  Methods in Natural Language Processing and Computational Natural Language
  Learning (EMNLP-CoNLL)}, page 12{\textendash}21.

\bibitem[{Si et~al.(2019)Si, Wang, Xu, and Roberts}]{si_enhancing_2019}
Yuqi Si, Jingqi Wang, Hua Xu, and Kirk Roberts. 2019.
\newblock \href {http://arxiv.org/abs/1902.08691} {Enhancing {Clinical}
  {Concept} {Extraction} with {Contextual} {Embedding}}.
\newblock \emph{arXiv:1902.08691 [cs]}.
\newblock ArXiv: 1902.08691.

\bibitem[{Smolensky(1986)}]{Smolensky1986}
P.~Smolensky. 1986.
\newblock \href {http://dl.acm.org/citation.cfm?id=104279.104290} {Parallel
  distributed processing: Explorations in the microstructure of cognition, vol.
  1}.
\newblock chapter Information Processing in Dynamical Systems: Foundations of
  Harmony Theory, pages 194--281. MIT Press, Cambridge, MA, USA.

\bibitem[{Smolensky(1990)}]{Smolensky1990TensorPV}
P.~Smolensky. 1990.
\newblock \href {https://doi.org/10.1016/0004-3702(90)90007-M} {Tensor product
  variable binding and the representation of symbolic structures in
  connectionist systems}.
\newblock \emph{Artif. Intell.}, 46(1-2):159--216.
\newblock \url{https://doi.org/10.1016/0004-3702(90)90007-M}.

\bibitem[{Smolensky et~al.(2016)Smolensky, Lee, He, tau Yih, Gao, and
  Deng}]{Smolensky2016BasicRW}
Paul Smolensky, Moontae Lee, Xiaodong He, Wen tau Yih, Jianfeng Gao, and
  Li~Deng. 2016.
\newblock Basic reasoning with tensor product representations.
\newblock \emph{CoRR}, abs/1601.02745.

\bibitem[{Stanovsky et~al.(2017)Stanovsky, Gruhl, and Mendes}]{Stanovsky2017}
Gabriel Stanovsky, Daniel Gruhl, and Pablo Mendes. 2017.
\newblock \href {https://www.aclweb.org/anthology/E17-1014} {Recognizing
  mentions of adverse drug reaction in social media using knowledge-infused
  recurrent models}.
\newblock In \emph{Proceedings of the 15th Conference of the {E}uropean Chapter
  of the Association for Computational Linguistics: Volume 1, Long Papers},
  pages 142--151, Valencia, Spain. Association for Computational Linguistics.

\bibitem[{Swayamdipta et~al.(2018)Swayamdipta, Thomson, Lee, Zettlemoyer, Dyer,
  and Smith}]{Swayamdipta2018}
Swabha Swayamdipta, Sam Thomson, Kenton Lee, Luke Zettlemoyer, Chris Dyer, and
  Noah~A. Smith. 2018.
\newblock \href {http://arxiv.org/abs/1808.10485} {Syntactic scaffolds for
  semantic structures}.
\newblock \emph{CoRR}, abs/1808.10485.

\bibitem[{Turian et~al.(2010)Turian, Ratinov, and Bengio}]{Ratinov2010}
Joseph Turian, Lev-Arie Ratinov, and Yoshua Bengio. 2010.
\newblock \href {https://www.aclweb.org/anthology/P10-1040} {Word
  representations: A simple and general method for semi-supervised learning}.
\newblock In \emph{Proceedings of the 48th Annual Meeting of the Association
  for Computational Linguistics}, pages 384--394, Uppsala, Sweden. Association
  for Computational Linguistics.

\bibitem[{Turney and Pantel(2010)}]{Turney2010}
Peter~D. Turney and Patrick Pantel. 2010.
\newblock \href {https://doi.org/10.1613/jair.2934} {{From frequency to
  meaning: Vector space models of semantics}}.
\newblock \emph{Journal of Artificial Intelligence Research}, 37:141--188.
\newblock \url{https://doi.org/10.1613/jair.2934}.

\bibitem[{Vaswani et~al.(2017)Vaswani, Shazeer, Parmar, Uszkoreit, Jones,
  Gomez, Kaiser, and Polosukhin}]{Vaswani2017}
Ashish Vaswani, Noam Shazeer, Niki Parmar, Jakob Uszkoreit, Llion Jones,
  Aidan~N Gomez, \L~ukasz Kaiser, and Illia Polosukhin. 2017.
\newblock \href
  {http://papers.nips.cc/paper/7181-attention-is-all-you-need.pdf} {Attention
  is all you need}.
\newblock In I.~Guyon, U.~V. Luxburg, S.~Bengio, H.~Wallach, R.~Fergus,
  S.~Vishwanathan, and R.~Garnett, editors, \emph{Advances in Neural
  Information Processing Systems 30}, pages 5998--6008. Curran Associates, Inc.

\bibitem[{Weston et~al.(2016)Weston, Bordes, Chopra, and
  Mikolov}]{Weston2016TowardsAQ}
Jason Weston, Antoine Bordes, Sumit Chopra, and Tomas Mikolov. 2016.
\newblock \href {https://arxiv.org/abs/1502.05698} {Towards ai-complete
  question answering: A set of prerequisite toy tasks}.
\newblock \emph{CoRR}, abs/1502.05698.

\bibitem[{Weston et~al.(2008)Weston, Ratle, and Collobert}]{Weston2008}
Jason Weston, Fr{\'e}d{\'e}ric Ratle, and Ronan Collobert. 2008.
\newblock \href {https://doi.org/10.1145/1390156.1390303} {Deep learning via
  semi-supervised embedding}.
\newblock In \emph{Proceedings of the 25th International Conference on Machine
  Learning}, ICML '08, pages 1168--1175, New York, NY, USA. ACM.
\newblock \url{https://doi.org/10.1145/1390156.1390303}.

\bibitem[{Wu et~al.(2016)Wu, Schuster, Chen, Le, Norouzi, Macherey, Krikun,
  Cao, Gao, Macherey, Klingner, Shah, Johnson, Liu, Kaiser, Gouws, Kato, Kudo,
  Kazawa, Stevens, Kurian, Patil, Wang, Young, Smith, Riesa, Rudnick, Vinyals,
  Corrado, Hughes, and Dean}]{wu_googles_2016}
Yonghui Wu, Mike Schuster, Zhifeng Chen, Quoc~V. Le, Mohammad Norouzi, Wolfgang
  Macherey, Maxim Krikun, Yuan Cao, Qin Gao, Klaus Macherey, Jeff Klingner,
  Apurva Shah, Melvin Johnson, Xiaobing Liu, Łukasz Kaiser, Stephan Gouws,
  Yoshikiyo Kato, Taku Kudo, Hideto Kazawa, Keith Stevens, George Kurian,
  Nishant Patil, Wei Wang, Cliff Young, Jason Smith, Jason Riesa, Alex Rudnick,
  Oriol Vinyals, Greg Corrado, Macduff Hughes, and Jeffrey Dean. 2016.
\newblock \href {http://arxiv.org/abs/1609.08144} {Google's {Neural} {Machine}
  {Translation} {System}: {Bridging} the {Gap} between {Human} and {Machine}
  {Translation}}.
\newblock \emph{arXiv:1609.08144 [cs]}.
\newblock ArXiv: 1609.08144.

\bibitem[{Zhang et~al.(2015)Zhang, Zhao, and LeCun}]{Zhang2015}
Xiang Zhang, Junbo~Jake Zhao, and Yann LeCun. 2015.
\newblock \href {http://arxiv.org/abs/1509.01626} {Character-level
  convolutional networks for text classification}.
\newblock \emph{CoRR}, abs/1509.01626.

\end{thebibliography}
\bibliographystyle{acl_natbib} 

\end{document}